# SIMULATED PERFORMANCE OF THE PRODUCTION TARGET FOR THE MUON G-2 EXPERIMENT*

D. Stratakis†, M. Convery, J. P. Morgan, D. Still, M. J. Syphers[1], Fermi National Accelerator Laboratory, Batavia IL, USA

V. Tishchenko, Brookhaven National Laboratory, Upton NY, USA

[1]also at Northern Illinois University, DeKalb IL, USA

*Abstract*

The Muon g-2 Experiment plans to use the Fermilab Recycler Ring for forming the proton bunches that hit its production target. The proposed scheme uses one RF system, 80 kV of 2.5 MHz RF. In order to avoid bunch rotations in a mismatched bucket, the 2.5 MHz is ramped adiabatically from 3 to 80 kV in 90 ms. In this study, the interaction of the primary proton beam with the production target for the Muon g-2 Experiment is numerically examined.

## INTRODUCTION

The Muon g-2 Experiment, at Fermilab [1], will measure the muon anomalous magnetic moment, $a_\mu$ to unprecedented precision: 0.14 parts per million. To perform the experiment, a polarized beam of positive muons is injected into a storage ring with a uniform magnetic field in the vertical direction. Since the positron direction from the weak muon decay is correlated with the spin of the muon, the precession frequency is measured by counting the rate of positrons above an energy threshold versus time. The g-2 value is then proportional to the precession frequency divided by the magnetic field of the storage ring.

A sequence of lines that are part of the Fermilab Muon Campus [2] have been designed in order to transport the highest possible quality beam to the Muon g-2 Experiment. An RF system in the Recycler separates a batch into 8 tighter bunches of $10^{12}$ protons each. These bunches are extracted and guided to a target station. The resulting pions, protons and muons are transported into the Delivery Ring (DR), where they make several revolutions. This will give enough time for all pions to decay to muons as well as provide enough separation in order to remove the protons with minimum losses. Finally, the muon beam is injected into a final beamline that terminates at the entrance of the storage ring of the Muon g-2 Experiment. In this paper, we will overview the performance of the production target.

## MUON PRODUCTION TARGET

The production target [3] station consists of five main devices: the pion production target, the lithium lens, a collimator, a pulsed magnet, and a beam dump (not depicted here). A schematic layout is shown in Fig. 1.

The new target design is made of a single cylinder of Inconel, with air blowing through a heat exchanger incorporated into the center shaft. A shell of beryllium provides a cover for the Inconel target, to reduce target oxidation and damage. Inconel was chosen as the best choice of target material because it can withstand higher stresses caused by the rapid beam heating. Immediately downstream of the target module is the Lithium Lens module. The lens is designed to focus a portion of the secondaries off of the target, greatly reducing their angular component. The distance between the target and lens, can be adjusted to match the diverging cone of secondary particles to the focal length of the Lithium Lens. For the Muon Campus scenario, the lithium lens acts as a conductor for a 116 kA current, producing a magnetic field gradient of 232 T/m within the lithium and is located 0.3 m downstream the target. The Lithium Lens has the advantage over conventional quadrupoles in that it focuses in both transverse planes and produces an extremely strong magnetic field. The following collimator is used to reduce heating and radiation damage to the Pulsed Magnet (PAMG), which is located immediately downstream of the Collimator. The Collimator is cylindrical in shape and made of copper, with a hole in the middle for the beam to pass through. The PMAG is a 3-degree pulsed dipole that is located downstream of the Collimator. Its purpose is to select 3.1 GeV/c secondaries and bend them into the M2 line. The dipole was designed specifically for the Target Vault and is a single-turn, radiation-hardened, water-cooled, 1.07 m long magnet with an aperture measuring 5.1 cm horizontally by 3.5 cm vertically.

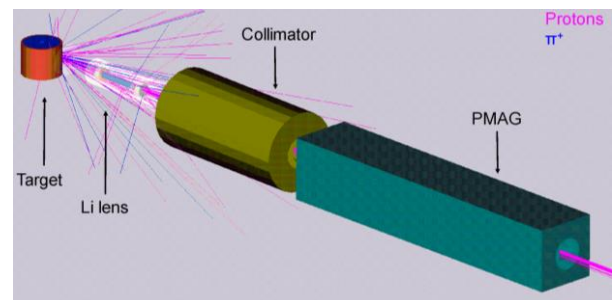

Figure 1: Schematic layout of the target area that is used to produce muons for the Muon g-2 Experiment. Note that the beam dump is not shown here.

## TARGET PERFORMANCE

The performance of the Muon Campus beamlines was simulated using G4Beamline [4]. G4beamline is an open-source particle simulation program that acts as a front-end for the Geant4 simulation package. While typical Geant4 simulations are written in C++, G4beamline programs are

---

written in a self-describing language. G4beamline is currently used in many muon projects, including the Mu2e experiment and the Muon Accelerator Program.

During the period when the Muon g-2 Experiment will take data, the Booster is expected to run with present intensities of $16 \times 10^{12}$ protons at 12 Hz average in two groups of 8 bursts at 100 Hz. There will be a group of two such bursts in a cycle of 1.4 s. The scheme for g-2 bunch formation [5] uses one RF system, 80 kV of 2.5 MHz. In order to avoid bunch rotations in a mismatched bucket, the 2.5 MHz is ramped adiabatically from 3 to 80 kV in 90 ms. Initially the bunches are injected from the Booster into matched 53 MHz buckets (80 kV of 53 MHz RF), then the 53 MHz voltage is turned off and the 2.5 MHz is turned on at 3 kV and then ramped to 80 kV. The resulting beam profile is illustrated in Fig. 2. Note that because the revolution time of muons in the storage ring is 149 ns, the longitudinal extent of the bunches should be no more than 120 ns. In addition, their 10 ms spacing allows enough time for muon decay and data acquisition in the storage ring.

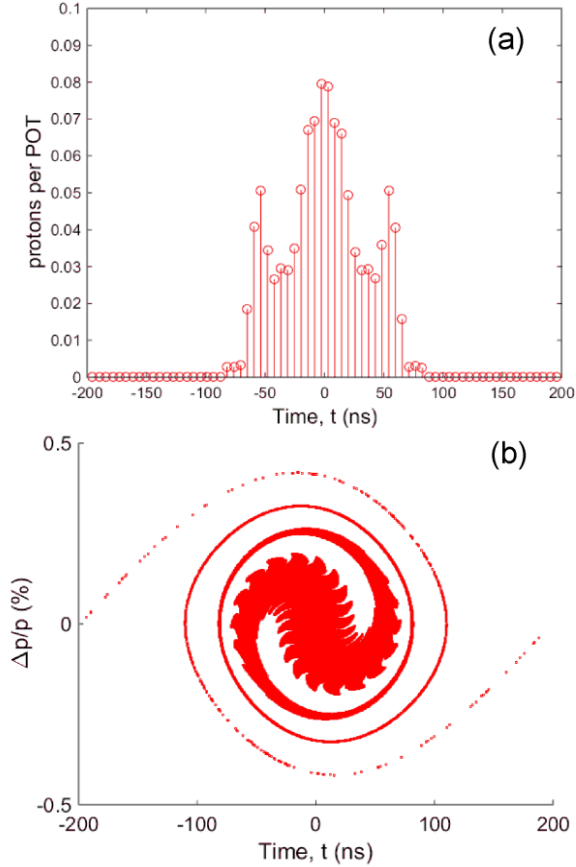

Figure 2: (a) Simulated profile of a proton bunch exiting the recycler, and (b) longitudinal phase space distribution centered at 8.89 GeV/c momentum.

Key beam parameters were optimized to maximize the downstream yield of secondary particles. For instance, it was found that the optimum proton beam spot size at the target is $\sigma_x = \sigma_y = 0.15$ mm. Figure 3 displays the secondary particles after the proton beam impacts the target. The population of particles in that location is: $1.83 \times 10^{-5}$ $\mu^+$, 0.725 protons, $2.72 \times 10^{-3}$ $\pi^+$, and $1.26 \times 10^{-4}$ $K^+$ per proton on target (POT). One interesting feature is that all secondary particles retain the same time profile as the initial proton pulse. Namely, 95% of the beam is concentrated within a 120 ns time span.

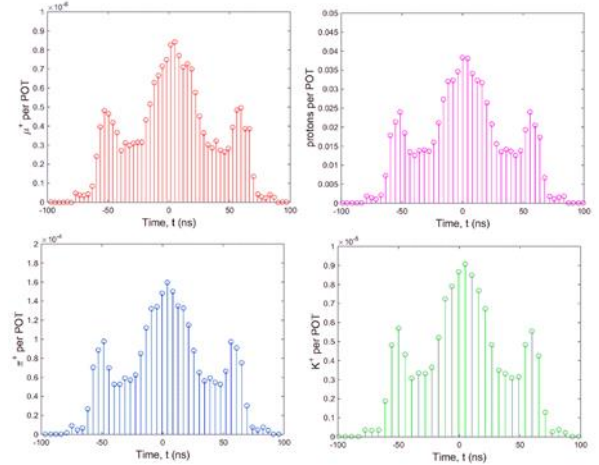

Figure 3: Longitudinal profiles of the secondary beam after it exits the Lithium lens (upstream of PMAG).

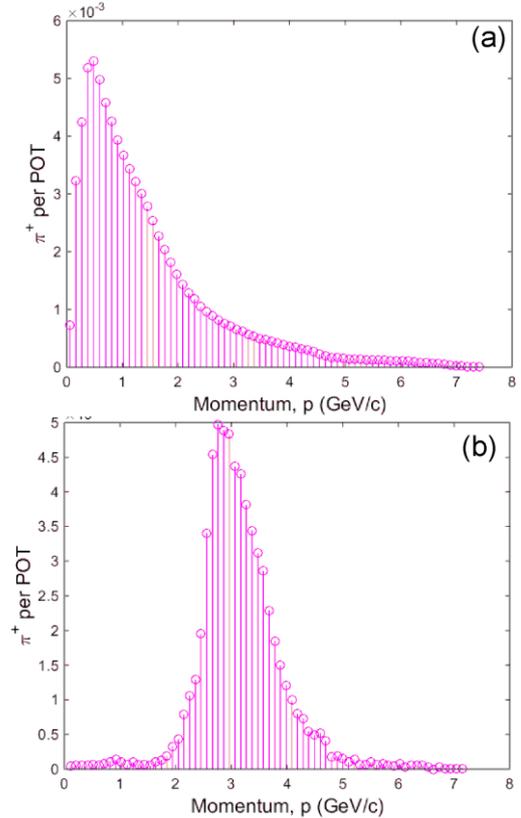

Figure 4: Momentum distribution (a) before PMAG and, (b) after the PMAG. PMAG selects particles near 3.1 GV/c.

Figure 4 displays the momentum distribution before and after the pulsed magnet. We can clearly see that the PMAG selects particles near the desired 3.1 GeV/c momentum for the Muon g-2 Experiment. Note that the distribution in Fig. 4(b) shows a slight asymmetry. This fact is not surprising since the vast majority of the pions entering the magnet have a momentum that is much less than the desired 3.1 GeV/c. Figure 5 displays the transverse angle of the selected pions from Fig. 4(b). Clearly, one can see that the pions with the most forward momentum (lowest angle with the beam axis) are around the momentum for which we are selecting. This demonstrates that the pulsed magnet behaves as expected and ensures that particles near 3.1 GeV/c will have the highest possible transmission in the downstream beamlines.

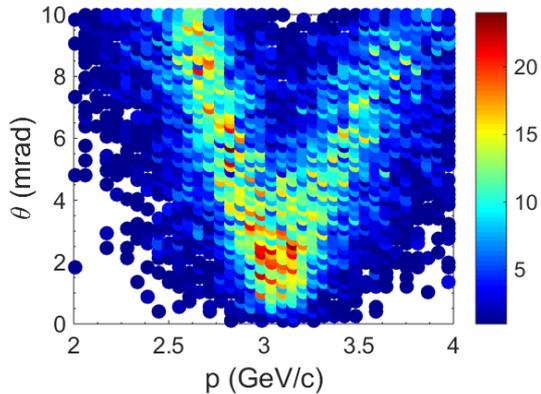

Figure 5: Pion momentum as function of the transverse angle after the beam passes the pulsed magnet. Particles near 3.1 GeV/c have the lowest possible transverse angle.

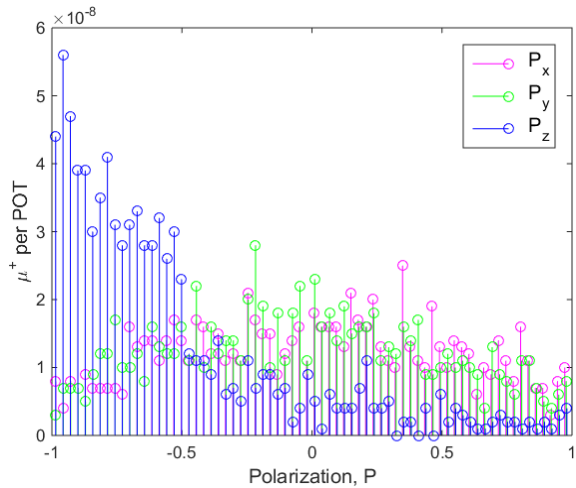

Figure 6: Vertical, horizontal and longitudinal components of polarization just upstream of PMAG. The average polarization in the longitudinal direction is 0.55. The average polarization in the horizontal and vertical planes is near zero.

One of the key aspects of the Muon g-2 Experiment is the delivery of a highly-polarized muon beam from pion decays. Along the M2 and M3 beamlines, the momentum acceptance is relatively narrow, ($\Delta p/p = \pm 2\%$) thereby producing highly-polarized muons. In contrast, the momentum distribution of pions in the target station is very broad, resulting in muons with a wide range of polarization. The distribution of polarization components of muons within ($\Delta p/p = \pm 2\%$) just upstream of PMAG is shown in Fig. 6. One can see that while the average polarization in the horizontal and vertical directions is near zero, the average polarization in the longitudinal direction is 0.55. Note that the polarization of muons arising from the decay of Kaons is excluded in this analysis.

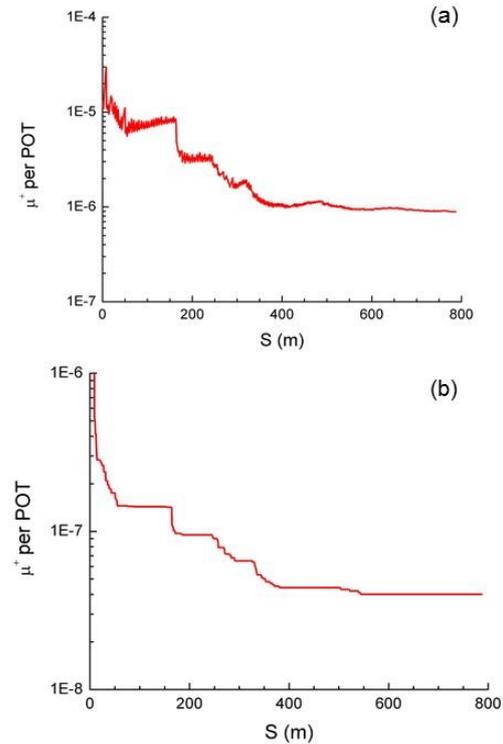

Figure 7: Propagation of muons along the M2, M3 lines and the Delivery Ring. Only one turn is assumed: (a) muons from pion decays after PMAG, and (b) muons created within the target area (upstream of PMAG).

Figure 7 shows the evolution of muons along the Muon Campus lines. Note that the entrance to the DR is at $S = 280$ m. Figure 7(a) shows the evolution of muons created from pion decays further downstream while Fig. 7(b) displays the evolution of "target produced" muons. Note that term "target produced muons" refers to those generated from pion decays in the area upstream PMAG. After one turn, the number of target produced muons is $4 \times 10^{-8}$ per POT while the muons from pion decays along the M2, M3 and DR is $9 \times 10^{-7}$ per POT. This suggests that among all muons after the first turn, only 4.5% are created upstream of PMAG. We conclude that the reduced polarization of the target muons is not expected to harm the overall polarization which is found to be >90% [6]. The authors are grateful to Raffaele Miceli for his help in setting up the G4Beamline model of the target.

## REFERENCES

[1] J. Grange et al., Muon (g-2) Technical Design Report, arXiv:1501.06858 (2015)
[2] muon.fnal.gov
[3] Fermilab Beams division. Antiproton Source Rookie Book.
[4] T. J. Roberts et al., in Proceedings of 2008 EPAC, (EPS-AG, Genoa, Italy, 2008), p. 2776
[5] I. Kourbanis, G Minus 2 Experiment Document 335-v1 (2012), https://gm2-docdb.fnal.gov/.
[6] D. Stratakis et al., Proceedings of IPAC 2016, p. 996 (2016